\documentstyle[12pt,epsf]{article}
\def\beq{\begin{equation}}
\def\eeq{\end{equation}}
\def\bea{\begin{eqnarray}}
\def\eea{\end{eqnarray}}
\def\bq{\begin{quote}}    
\def\eq{\end{quote}}

\def\bq{\begin{quote}}
\def\eq{\end{quote}}

\def\npb#1#2#3{\mbox{Nucl. Phys. {\bf B#1} (#2) #3}}
\def\plb#1#2#3{\mbox{Phys. Lett. {\bf B#1} (#2) #3}}
\def\prd#1#2#3{\mbox{Phys. Rev. {\bf D#1} (#2) #3}}
\def\prl#1#2#3{\mbox{Phys. Rev. Lett. {\bf #1} (#2) #3}}
\def\prc#1#2#3{\mbox{Phys. Rep. {\bf #1} (#2) #3}}

\def\ijmp#1#2#3{\mbox{Int. J. Mod. Phys. {\bf A#1} (#2) #3}}

\begin{document}

\baselineskip 24pt

\newcommand{\sheptitle}
{F-term Hybrid Inflation in Effective Supergravity Theories}

\newcommand{\shepauthor}
{M. Bastero-Gil$^\ast$ and S. F. King$^{\dagger}$ }

\newcommand{\shepaddress}
{$^\ast$Department of Physics and Astronomy,
University of Southampton, \\ Southampton, SO17 1BJ, U.K.\\
$^\dagger$ Theory Division, CERN, CH-1211 Geneva 23, Switzerland}

\newcommand{\shepabstract}
{We show that a particular class of
effective low energy supergravity theories motivated by
string theory can provide a promising framework for models
of hybrid inflation in which the potential energy which drives
inflation originates from the F-term of the effective supergravity
theory. In the class of models considered 
the inflaton is protected from receiving mass during inflation by
a generalisation of the Heisenberg symmetry present in no-scale 
supergravity models. 
The potential during inflation takes the
positive definite form  
$V\sim |F_S|^2 + |F_T|^2 -3$, which allows the possibility that $V\ll
m_{3/2}^2 M_P^2$ through the cancellation of the positive dilaton and
moduli contribution against the negative term. 
We discuss a toy example where this is realised, then describe 
the application of this result to realistic models focusing on a
particular example in which the
$\mu$ problem and the strong CP-problem are addressed.}

\begin{titlepage}
\begin{flushright}
CERN-TH/98-190\\
SHEP 98-10 \\
hep-ph/9806477\\
\end{flushright}
\vspace{.1in}
\begin{center}
{\large{\bf \sheptitle}}
\bigskip \\ \shepauthor \\ \mbox{} \\ {\it \shepaddress} \\ \vspace{.3in}
{\bf Abstract} \bigskip \end{center} \setcounter{page}{0}
\shepabstract
\begin{flushleft}
CERN-TH/98-190\\
\today
\end{flushleft}
\end{titlepage}

\section{Introduction}

Due to its intrinsic elegance, inflation \cite{inflation} has become the 
almost universally accepted dogma for 
accounting for the flatness and homogeneity of the universe.
There are various classes of inflation that have been proposed,
but possibly the most successful, and certainly one of the most
popular versions these days is hybrid inflation \cite{hybrid,copeland}. 
In hybrid inflation,
there are (at least) two fields at work: the slowly rolling inflaton
field $\phi$, and a second field which we shall call $N$ whose 
value is held at zero during inflation and whose role is to end inflation
by developing a non-zero vacuum expectation value (VEV) when $\phi$
passes a certain critical value $\phi_c$ during its slow roll.
With $N=0$, the potential along the $\phi$ direction is approximately flat,
with the flatness lifted by a $\phi$ mass which must be small enough
to satisfy the slow-roll conditions for inflation. 

The natural framework for hybrid inflation is supersymmetry.
SUSY can naturally provide flat directions along which the inflaton
can roll, and additionally ensures that the scalar inflaton mass
does not have quadratic divergences. The natural size of the
inflaton mass in SUSY is of order the SUSY breaking scale, and the slow roll
conditions and COBE constraints \cite{deltah} then determine the
height of the potential 
during inflation $V(0)^{1/4}$ to be some intermediate scale below the
grand unification (GUT) scale. The precise value of $V(0)$ is
model-dependent, however the lower the height of the potential, the
flatter the inflaton potential has to be and smaller the inflaton mass.

If one accepts the above framework, one is driven almost inevitably
to supergravity (SUGRA).
The origin of the vacuum energy $V(0)$ which drives inflation 
can only be properly understood within a framework which allows the
possibility for the
potential energy to settle to zero at the
global minimum, and hence lead to an acceptable cosmological constant, and
this implies SUGRA. In this way we are led to consider SUGRA models
in which the fields are displaced from their global minimum values
during inflation. 
The potential in SUGRA is given by:
\beq
V = e^G \left[ G_i (G^i_{\bar{j}})^{-1} G^{\bar{j}} - 3 \right] + |D|^2 \,,
\label{V}
\eeq
in the usual notation, i.e. natural units, 
with subscripts $i$ ($\bar{i}$)
referring to partial differentiation with respect to the generic 
field $\phi_i$ ($\phi_i^\ast$).
We have written the D-terms very schematically as $|D|^2$.
The Kahler function $G$ is:
\beq
G=K +\ln |W|^2 \,,
\label{G}
\eeq
where the Kahler potential $K$ is a real function of generic fields
$\phi , \phi^\ast$ and the superpotential $W$ is an analytic (holomorphic)
function of $\phi$ only. The Kahler metric is 
$G_i^{\bar{j}}$ and its inverse satisfies 
$(G^i_{\bar{j}})^{-1}G_k^{\bar{j}}=\delta^i_k$.
The SUGRA F-terms are:
\beq
F^i=e^{G/2}(G^i_{\bar{j}})^{-1} G^{\bar{j}} \,,
\label{F}
\eeq
whose non-zero value signals SUSY breaking, with a gravitino mass
\beq
m_{3/2}^2=e^{G}= e^K |W|^2\,.
\label{m3/2}
\eeq

Using the preceding results we can schematically write the potential in
SUGRA as:
\beq
V=|F|^2 + |D|^2 -3m_{3/2}^2 \tilde{M}_P^2 \,,
\label{Vschematic}
\eeq
where we have put back the reduced Planck mass $\tilde{M}_P$.
Eq. (\ref{Vschematic}) shows that there are two possible sources
for the positive 
vacuum energy $V(0)$ which drives inflation: the F-term or the
D-term. The negative term also allows for eventual cancellation of the
potential energy, and is the main motivation for considering SUGRA.
Assuming that the D-terms are zero and the F-terms
of the same order of magnitude
during inflation as they are at the end of inflation, 
we have\footnote{After inflation, since $V=0$ we must have 
$m_{3/2}=\frac{F}{\sqrt{3}\tilde{M_P}}$ at the global minimum, which
is the Deser-Zumino relation. 
It is hard to see how this relation could be so badly violated
during inflation so that typically one expects that during inflation
$m_{3/2}\sim \frac{F}{\tilde{M_P}}$.}
\begin{equation}
 V(0) = |F|^2 -3m_{3/2}^2{\tilde{M_P}}^2
\sim m^2_{3/2} {\tilde{M_P}}^2 \sim (10^{11}\, GeV)^4  \label{pot} \,.
\end{equation}
Here the gravitino mass is assumed to be of order of 1 TeV.
Now if we make the further assumption that
the inflaton $\phi$ mass is of order $m_{3/2}$, this line of 
reasoning leads to a violation of the slow roll condition,
\begin{equation}
|\eta | ={\tilde{M_P}}^2 \left( \frac{V''}{V} \right) \ll 1 \,. 
\end{equation}
In fact from Eq. (\ref{pot})
we predict $|\eta| \sim 1$.
This is the so-called $\eta$ problem \cite{eta}.
To overcome it, we must relax one or more of the assumptions in the
chain of logic that led to it. For example, one can imagine that
both during and after inflation the F-term and the -3 term in 
Eq. (\ref{Vschematic}) exactly cancel, and that the energy which drives
inflation originates from the D-term \cite{D}. Then $|D|^2$ is allowed to take
a higher value than that in Eq. (\ref{pot}), providing it cancels
to zero at the end of inflation; indeed the problem 
in D-term inflation is rather one of keeping
the potential small enough compared to the string
scale\footnote{D-term inflation 
may be more complicated than originally thought due to the unavoidable
contribution of the dilaton in the F-term sector to SUSY breaking masses
\cite{dilaton}.} \cite{string}. We will not address this problem in
this paper, and from here on we will assume that the D-terms vanish
during inflation. 

One interesting loop-hole in the above argument
which we would like to pursue in the present paper is the possibility that
the inflaton mass is in fact much smaller than $m_{3/2}$ during inflation.
For example it is known that in no-scale SUGRA theories, the soft scalar
masses are zero (up to radiative corrections)
even in the presence of a non-zero gravitino mass \cite{noscale}.
These no-scale results apply for all values of the fields, including
during inflation when the fields are
away from the global minimum. In fact the general conditions under which
the inflaton maintains a zero mass during inflation have been formalised
in terms of a Heisenberg symmetry of the Kahler potential
\cite{heisenberg}. However as shall become clear later
the approach that we follow is more general than the Heisenberg symmetry.
In fact the basic requirements of a successful F-term theory of inflation
will be seen to be a no-scale assumption for the inflaton, together with
other conditions which guarantee that
it remains massless at tree-level during inflation, plus the requirement
that the inflaton couples sufficiently weakly to other sectors so that
the radiative corrections to its mass are very small. The requirement of
small inflaton mass, combined with the COBE normalisation, imply that the
height of the potential during hybrid inflation must be lower
than the usual value $10^{11}$ GeV, and we show how the no-scale
structure allows this possibility.
In fact we shall give an explicit example of our approach 
where the SUSY breaking sector, the height of the potential and the
inflation sector are all specified.
The example involves an ultra-light inflaton whose mass after radiative
corrections is in the eV range. For such an ultra-light inflaton, the
COBE normalisation demands that the potential during inflation be very
low, of order $10^8$ GeV.

\section{Effective No Scale Supergravity Models From String Theory.}

The simplest example of a no-scale SUGRA model consists of a Kahler
potential
\beq
K=-3\ln(T+T^\ast- \phi^\ast \phi) \,,
\label{no-scaleK}
\eeq
and a superpotential which is independent of the field $T$,
\beq
W=W(\phi)\,.
\label{no-scaleW}
\eeq
When the potential is constructed using Eqs. (\ref{V}), (\ref{G}) one
finds firstly that certain cross-terms involving derivatives
of the superpotential like $W_{\phi}$ cancel with each other, and secondly that
terms involving derivatives of the Kahler potential cancel with the -3.
The only term which survives is:
\beq
V = e^G \frac{\rho}{3|W|^2}|W_{\phi}|^2=\frac{1}{3\rho^2}|W_\phi|^2 \,,
\label{Vnoscale}
\eeq
where we have defined for convenience:
\beq
\rho \equiv T+T^\ast- \phi^\ast \phi \,.
\label{rho}
\eeq
The potential is minimised along the flat direction 
$W_\phi=0$, and in fact is identically zero for all values of $\rho$.
Supersymmetry is broken by the non-zero value of the F-term 
$F_T=-\rho e^{G/2}$ but neither it nor the gravitino mass
$m_{3/2}=e^{G/2}$ is determined since $\rho$ is not fixed at tree-level.
However, assuming $\rho$ is fixed by radiative/non-perturbative 
corrections, we  
conclude that supersymmetry is broken, the gravitino has a mass,
but the fields $\phi$ remain massless.
The masslessness of the inflaton may be associated with the so called
Heisenberg symmetry \cite{heisenberg} which is more or less equivalent to 
the observation that the fields $T,T^\ast$ only ever appear
in the Kahler function in the combination $\rho$, and so the
degrees of freedom may be regarded as $\rho = \rho^\ast$ and $\phi$,
and the masslessness of $\phi$
is guaranteed for any Kahler potential of the form $K_0(\rho)$.
On the other hand, we emphasise that if
one allows a superpotential with a non-trivial
$T$ dependence then extra terms appear in the potential proportional
to $W_T$ which invalidate these results. 


Although the no-scale SUGRA mechanism looks quite specialised,
it turns out that it fits in quite well within 4d effective
string theories. For example, in orbifold constructions
\cite{orbifold} the tree-level Kahler potential is given by,
\beq
K=-\ln(S+S^\ast)-3 \ln(T+T^\ast)
+\sum_{a} (T+T^\ast)^{n_a} C_a^\ast C_a \,,
\label{Kgen}
\eeq
where in the usual notation, $S$ is the dilaton, $T$ is 
an over-all modulus and $C_a$  observable (matter)
superfields with modular weights $n_a$. Untwisted sector superfields
have $n_a=-1$; fields belonging to the twisted sector 
without oscillators have usually
$n_a=-2$, and those with oscillators usually have $n_a\leq -3$,
but in general any integer $n_a\leq 0$ is possible for twisted fields.
The assumption of an over-all modulus $T$ is a simplified one and 
generically there are at least three $T_i$ moduli fields,
the over-all modulus case corresponding to the limit
$T_1=T_2=T_3$. Nevertheless, one would expect the values of the moduli fields
not to be very different, and approach the simpler over-all modulus
situation which we will consider in this paper. The differences that
may appear in the multimoduli case are not relevant for the discussion
below.

In dealing with inflation we shall assume that the
inflaton $\phi$ is an untwisted field and distinguish it
from all the other matter fields by writing
explicitly $\phi \equiv C_0$. Then the Kahler potential for the
modulus and the inflaton is virtually identical to the non-scale SUGRA
Kahler potential in Eq. (\ref{no-scaleK}), and we can rewrite $K$ as, 
\beq
K=-\ln(S+S^\ast)-3\ln(T+T^\ast- \phi^\ast
\phi)+\sum_{a>0}(T+T^\ast)^{n_a} C_a^\ast C_a \,, 
\label{K}
\eeq
and define $\rho= T+T^\ast- \phi^\ast \phi$ as done before. 
The superpotential (which by assumption is independent of $T$)
may be written as:
\beq
W=\hat{W}(S)+\tilde{W}(\phi,C_a) \,,
\label{W}
\eeq
where for simplicity we have assumed that 
$\tilde{W}(\phi ,C_a)$ has no $S$ dependence,
so that $\hat{W}$ is the superpotential in the ``hidden'' sector,
and $\tilde{W}$ is the superpotential in the ``observable'' sector.

To proceed further with the discussion we need to specify the
superpotential $W=\hat{W}+\tilde{W}$. 
This is  highly model dependent, and in particular
the dilaton superpotential $\hat{W}(S)$, which may be
responsible for SUSY breaking, is not predicted by string theory. 
In general both $K$ and $W$ may receive 
corrections related to the problem of dilaton stabilisation
\cite{dilaton0,dilaton1}, and 
different models have been proposed in the recent literature.
For example, a superpotential $\hat{W}(S)$ can be generated in such a way
that the dilaton 
is stabilised but $F_S=0$, i.e., with zero potential, both during and after
inflation, as recently discussed by Riotto and one of us \cite{KR}.
Here we are most interested in a situation with a non-zero $F_S$, and
for that purpose we take a model of gaugino condensation in which the
dilaton is stabilised by non perturbative corrections to its Kahler potential
\cite{dilaton1}. Following Ref. \cite{dilaton}  we take the particular
ansatz:
\bea
K(S)&=& - \ln (S+S^\ast) + \hat{K}_{np}(S) \,, \\
\hat{K}_{np}(S)&=& -\frac{2 s_0}{S+S^\ast} + \frac{b + 4 s_0^2}{6 (S+S^\ast)^2}
\label{Kdilanp} \,,
\eea
where $b$ and $s_0$ are non-negative constants; we also consider  the
dilaton superpotential:
\beq
\hat{W}(S)= \Lambda^3 e^{-S/b_0}  \label{WS}\,,
\eeq
with $\Lambda$ an unspecified scale for the time being. This effective
scale will be later fixed by the requirement of having $m_{3/2}\approx
1 \, TeV$ after inflation ends. 
The potential 
in the $S$ direction has a minimum at some value $ReS < s_0$, and
an exponentially decaying tail for $ReS>s_0$. The constant $b$ in the
non-pertubative Kahler potential controls the height of the barrier
between the minima and the exponentially decaying part, being the
barrier as larger as smaller is $b$.       

Including the dilaton contribution in the potential in Eq. (\ref{Vnoscale}), 
the potential in the $\rho$
direction goes simply as $\propto 1/\rho^3$. This would imply that
$\rho \rightarrow \infty$, unless the potential vanishes, and no period
of inflation be allowed. Therefore we must appeal to some
additional non-perturbative corrections to the no-scale Kahler
potential in order to stabilise the value of $\rho$. As a specific
example to model the non-perturbative contribution we take:
\beq
K_{np}(\rho)= \frac{\beta}{\rho^3} \,,
\label{betabit}
\eeq
with $\beta$ a positive constant.  

Joining together all the different pieces, Eq. (\ref{K}) is replaced by:
\beq
K= -3\ln(\rho) + \frac{\beta}{\rho^3} - \ln (S+S^\ast) + \hat{K}_{np}(S)
+\sum_{a>0}(T+T^\ast)^{n_a} C_a^\ast C_a   \label{Kfinal}\,.
\eeq
Note however that the theory in Eqs. (\ref{W}), (\ref{WS}) and
(\ref{Kfinal})  
does not have the Heisenberg symmetry since
the fields $T,T^\ast$  do not exclusively appear in the combination
of Eq. (\ref{rho}). However, since we are only interested in preserving the
masslessness of the inflaton {\it during inflation}, all we require is that
the remaining matter fields with non-zero modular weight
are switched off during inflation,
or at least do not contribute to the potential during inflation.
This can be seen as a constraint on the superpotential for such
matter fields rather than on the Kahler potential. This means that,
apart from the inflaton, the matter fields with non-zero modular weight
have to satisfy the conditions $C_a=W_a=0$
in the false vacuum of the theory where inflation is taking place,
even if they do not couple to the inflaton. 
Notice that for the field $N$ in hybrid inflation, with a typical
hybrid superpotential 
\beq
\tilde{W}(\phi,N)= -k \phi N^2 \,, 
\eeq
these conditions are inmediately fulfilled and thus its Kahler
potential may still be  
arbitrary.  We shall assume later that the singlet $N$ also belongs to
the untwisted sector, entering in the definition of $\rho=T+T^\ast -
\phi^\ast \phi - N^\ast N$. As far as inflation is concerned, this will
not make any difference. Note that $\tilde{W}$ is zero during
inflation and therefore the superpotential in Eq. (\ref{W}) is given
by the dilaton superpotential $\hat{W}$.   

The conditions that we propose
to protect the flatness of the inflaton potential, Heisenberg symmetry
and stabilising $\rho$ through non perturbative contributions,  should
be compared  to other approaches. 
In \cite{stewart} the requirement is that either the field $C_a$
or its derivative $W_a$ to be zero, with the additional
constraint that the superpotential during inflation must
vanish. This condition ensures that $dV/d\rho$ vanishes identically,
and $\rho$ is a flat direction. In such a situation the false vacuum
energy would be dominated 
by some non-zero $W_a$. By contrast
in our present scenario we can relax the
assumption of having a zero superpotential by the specific choice of the
inflaton Kahler potential we consider in order to stabilise $\rho$ at
some fixed value. In \cite{copeland}
a multimoduli Kahler potential is considered
with $all$ matter fields having modular weights -1 with
respect to one of the moduli $T_i$. With such a restriction they are
led to conclude that the
over-all modulus case is not viable for inflation, because in this case 
there would be no possible source to generate the potential energy
without giving a large mass to the inflaton. In our approach
this problem is
circumvented by the inclusion of the dilaton contribution. The
multimoduli Kahler potential 
in the context of inflation has also been recently studied in
\cite{lyth}.

\section{The Height of the SUGRA Inflation Potential.}

We now turn to the question of how to achieve a height of the potential
below the typical SUGRA value during inflation, when the
only relevant fields are $\rho$ and $S$. The potential obtained from
Eqs. (\ref{V}) and (\ref{G}) is given by, 
\beq
V(\rho,S)=e^K |W|^2 \left( \frac{K'^2}{K''} - 3 + \frac{1}{K_{SS}} \left| K_S
+ \frac{W_S}{W} \right|^2 \right) \,,
\label{Vrhos}
\eeq
where a subscript $S$ denotes derivative with respect to the
dilaton, primes represent derivation with respect to $\rho$; the
Khaler potential is given in Eq. (\ref{Kfinal}), $W$ is given in
Eq. (\ref{W}) with $\tilde{W}=0$ and then $W_S/W=-1/b_0$.  
Both fields, $\rho$ and $S$, will have a minimum at some value
$\rho_0$ and $S_0$ respectively, and they will contribute with a positive
term to the potential. Now, the size of the cosmological constant
during inflation will depend on the degree of fine-tuning we want to
impose on the cancellation of this contribution against the negative
term $-3 m^2_{3/2} \tilde{M}_P^2$. As discussed in the introduction,
we are interested in the situation where they almost cancel, i.e.,
\beq
V(\rho_0,S_0) = \epsilon m^2_{3/2} \tilde{M}_P^2\,,\;\; \epsilon \ll 1
\label{Vep} \,.
\eeq

\begin{figure}[t]
\epsfxsize=10cm
\epsfysize=10cm
\hfil \epsfbox{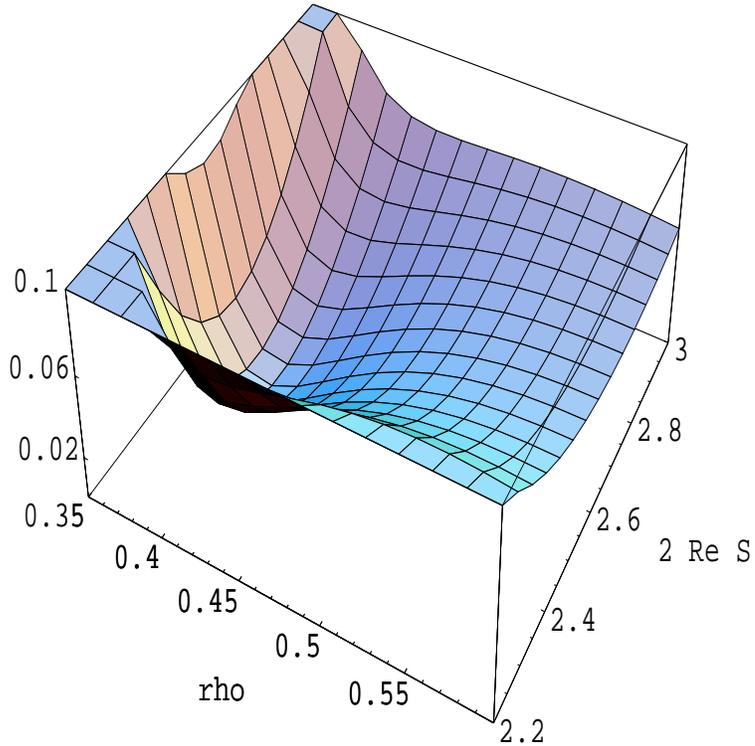} \hfil 
\caption{The potential $V(S,\rho)$ in units of $m_{3/2}^2
\tilde{M}_P^2$. $b=1$, $s_0=4$, $\beta=1/32$.} 
\end{figure}

The procedure is straighforward: we
just minimise the potential with the additional constraint
Eq. (\ref{Vep}). We find the three set of equations:
\bea
 \frac{K'^2}{K''} -3 + \frac{1}{K_{SS}} \left| K_S-
\frac{1}{b_0}\right|^2 &=& \epsilon \,,\\
 \frac{K' K'''}{K''^2}-2 &=& \epsilon \,, \label{minrho}\\
\frac{K_{SSS}}{K_{SS}^2}(K_S-\frac{1}{b_0})-2&=& \epsilon \,, \label{minS} 
\eea
which can be solved for $b_0$, $S$ and $\rho$, for given values of
$\beta$, $b$ and $s_0$. The solution for $\rho$ can be found
analytically as a series expansion in the parameter $\epsilon$, and is
given by: 
\bea
\rho_0 &=& (2 \beta)^{1/3} ( 1 + \epsilon + \epsilon^2 +\ldots )
\label{rho0}\,, \\
\frac{K'^2}{K''} -3 &=& -\frac{3}{4}( 1 - 3 \epsilon^2 + \ldots )
\,.
\eea
The values of
$b_0$ and $S_0$ (value at the minima)  are searched numerically. We notice that
neither $b_0$ nor $S_0$ depend on the parameter $\beta$. 
As an example, in Fig. (1) we have plotted the
potential $V(\rho,S)$, which
clearly shows the presence of a minimum, for the choice $b=1$,
$\beta=1/32$ and $s_0=4$ . In Fig. (2) we have plot the
projections $V(\rho_0,S)$ and $V(\rho,S_0)$ respectively, and in
Fig. (3) the computed values of $b_0$ versus the parameter $s_0$.  
The minimum of the potential is not exactly at zero, but at $\epsilon \simeq
10^{-10}$. This implies a potential scale of order
$10^8\,GeV$ during inflation if the gravitino mass is order
$1\,TeV$. Such a low value for $\epsilon$ can be 
achieved only by fine-tuning  $b_0$. 

Once we have found the value $\rho_0$ for which the condition
$dV/d\rho=0$ is fulfilled, it is not difficult to check that the
inflaton will remain massless. The mass matrix for
the fields ($T,\, \phi$) during inflation is given by:
\beq
{\cal M }^2(T,\phi) = \left( \begin{tabular}{cc} 
                            $V''$  &  $-\phi V''$ \\
                           $-\phi^\ast V''$ & $|\phi|^2 V''$
 	                   \end{tabular} \right)\,,
\eeq 
with $V''= d^2 V/d\rho^2$ at $\rho_0$. It is clear that this matrix
has a zero eigenvalue which corresponds to the massless inflaton, that
for simplicity we will call $\phi$. In
other words as the field $\phi$ rolls the moduli $T$ will adjust
itself in such a way that the overall potential remains flat.  

\begin{figure}[t]
\hspace{-.7cm}
\begin{tabular}{cc}
\epsfxsize=7.5cm
\epsfysize=8cm
\epsfbox{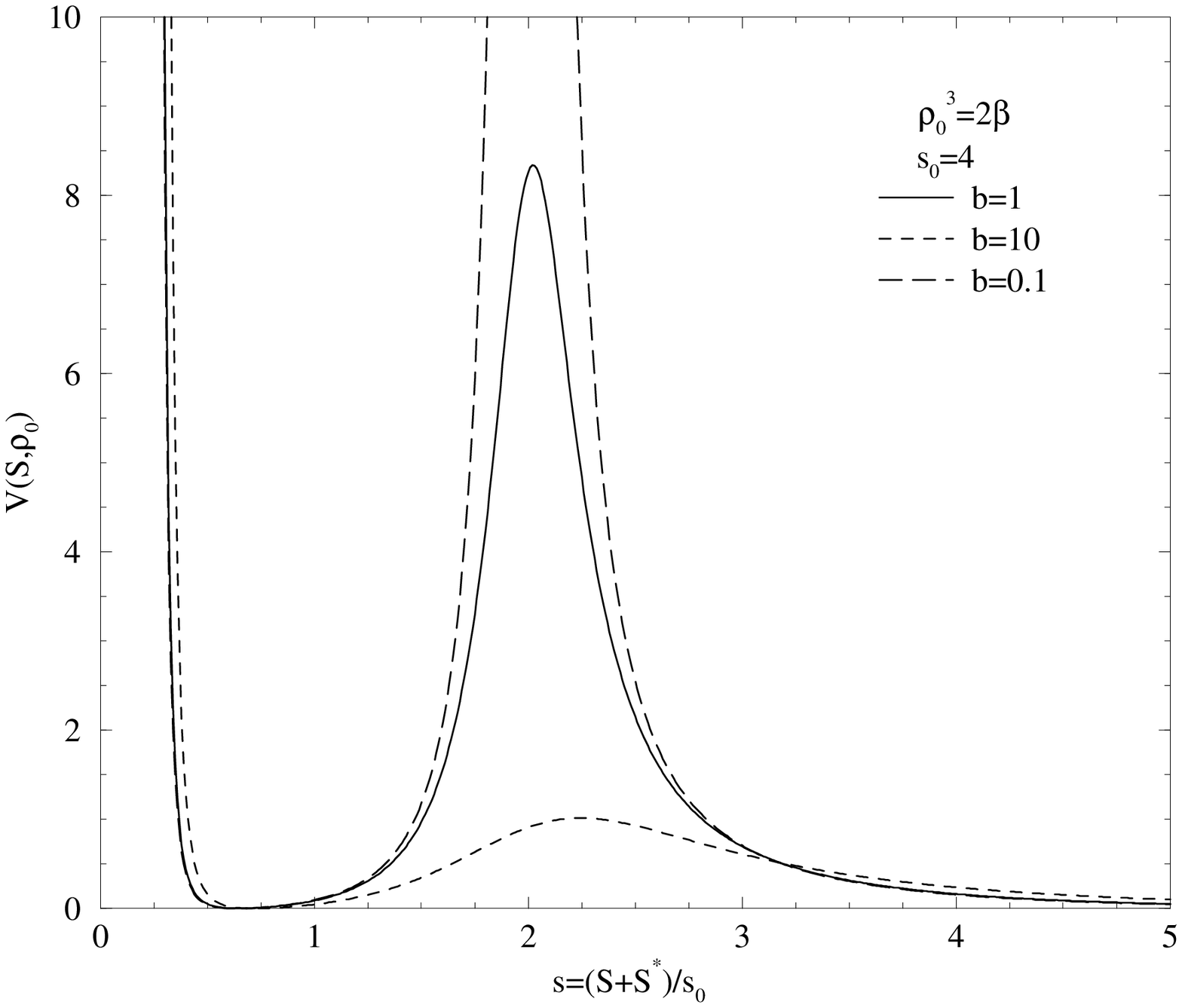} 
&
\epsfxsize=7.5cm
\epsfysize=8cm
\epsfbox{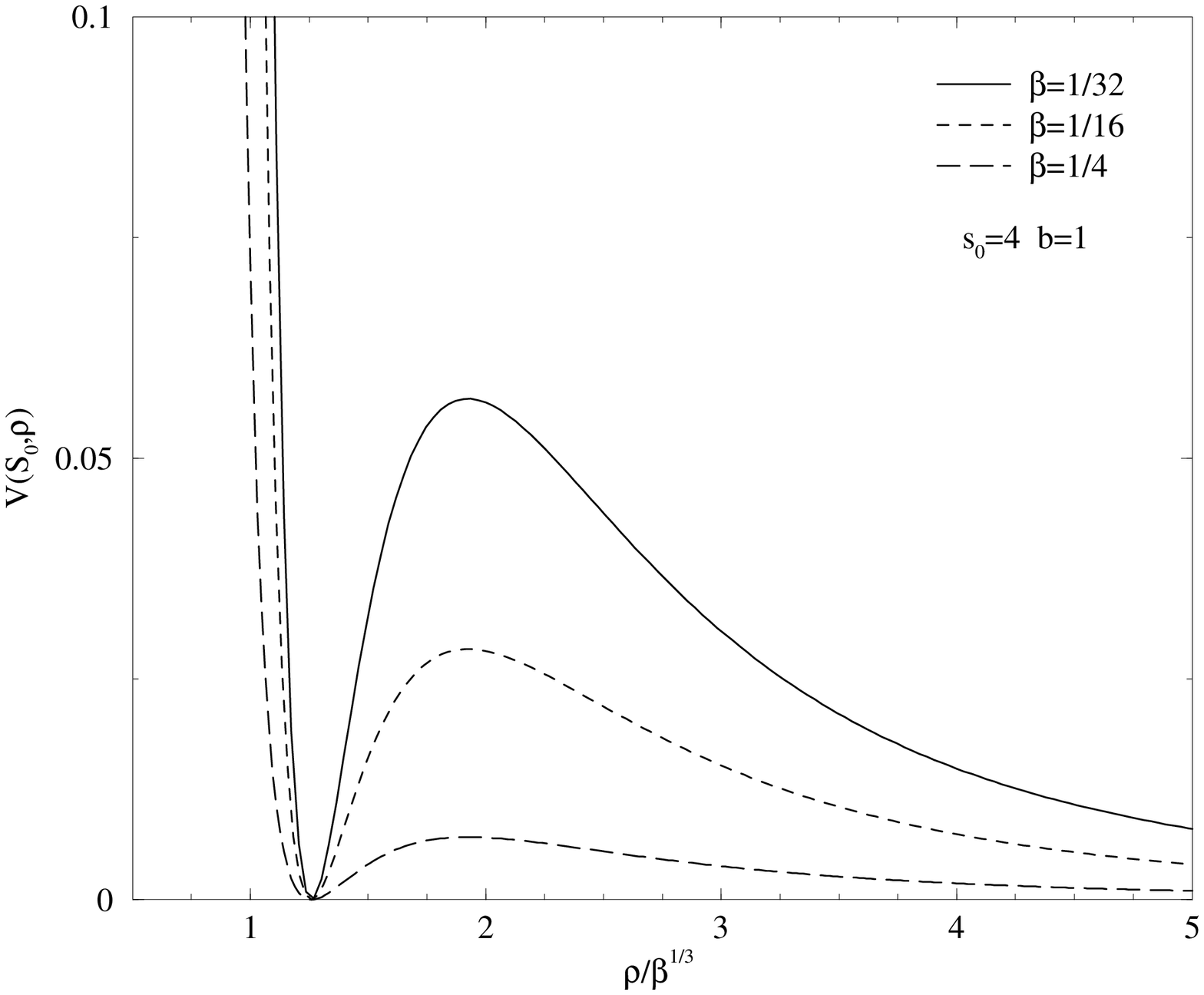} 
\end{tabular}
\hfill
\caption{(a) The potential $V(S,\rho_0)$, for
$s_0=4$, and different values of $b$.   
(b) The potential $V(S_0,\rho)$, for
$s_0=4$, $b=1$ and different values of $\beta$. The potential is given
in units of $m_{3/2}^2 \tilde{M}_P^2$. Note that here we have used
scaled variables, whereas in Fig. (1) we used unscaled ones.  }
\end{figure}

\begin{figure}[h]
\epsfxsize=10cm
\epsfysize=10cm
\hfil \epsfbox{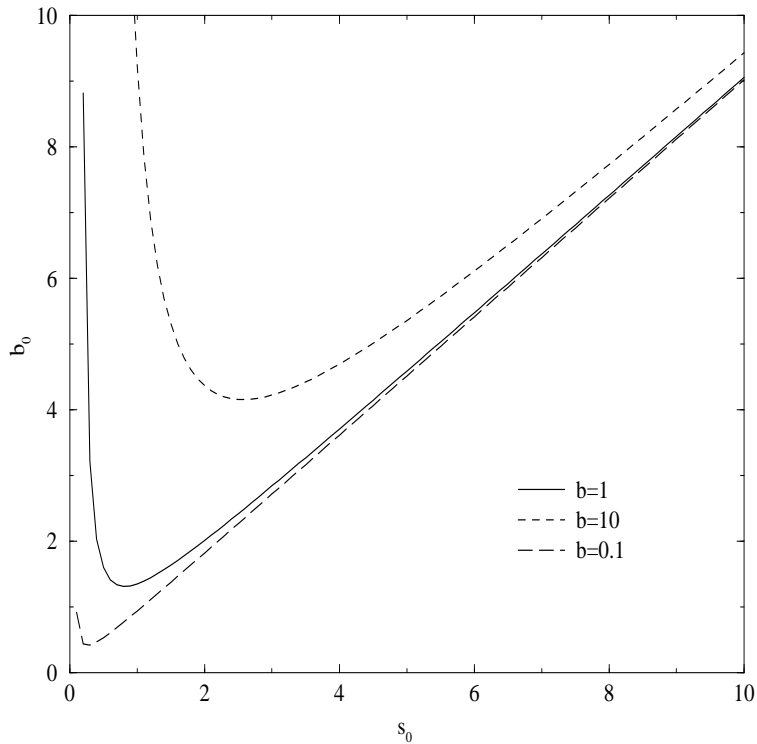} \hfil 
\caption{Computed values of $b_0$, with the condition $\epsilon \ll 1$.}
\end{figure}

\section{A Next-to-Minimal Supergravity Model of Hybrid Inflation.}

Finally we  give an example of a model of hybrid inflation
where these results may apply, namely the next-to-minimal 
supersymmetric model (NMSSM) of hybrid inflation which was recently
proposed \cite{NMSSM}. This model was proposed in the context of global
SUSY, where the origin of the (small)
vacuum energy which drives inflation
was not explained, the mechanism of supersymmetry breaking was
not discussed, the $\eta$ problem was not resolved, and the smallness
of the inflaton mass was simply assumed to be due to some
unspecified no-scale SUGRA model. We are now in a position to address
all these issues in the light of the preceding discussion.

We begin with a brief resume of the global SUSY model.
The model is based on the superpotential:
\beq
\tilde{W}=\lambda N H_1 H_2-k\phi N^2 \,,
\label{WNMSSM}
\eeq
with the fields $\phi , N$ being gauge singlets, and the first
term coupling the singlet $N$ to the MSSM Higgs doublets
$H_1 , H_2$. Since the VEV of $N$ generates the effective $\mu$ mass
term coupling the two Higgs doublets, we require that
$\lambda <N> \sim 1 $ TeV as in the well known particle physics
NMSSM model. The Higgs doublets develop electroweak
VEVs, much smaller than $<N>$, and they  may be ignored in the analysis.
Therefore, the potential for the real components of the complex
singlets reads,
\beq
V(\phi,N)= k^2 N^4 + (m^2_N-2 k A_k \phi + 4 k^2 \phi^2) N^2 +
m_{\phi}^2 \phi^2 \, \label{Vnphi},
\eeq
where the soft parameters above occur in the soft SUSY breaking potential
$V_{soft}=m_N N^2-A_k k\phi N^2$, but $m_{\phi}^2$ owes its origin to
radiative corrections to the potential controlled by the small
coupling $k$. We have fixed our convention of signs and phases such
that $k A_k >0$ but appears with a negative 
sign in the potential.  This negative contribution is neccesary not only
for the purpose of inflation but in order to get electroweak symmetry
breaking afterwards. 

For large values of the field $\phi$ the effective $N$ mass is
positive and during inflation the field $N=0$; $\phi$ slowly rolls until it
reaches a critical value:
\beq
 \phi_c^{\pm}=\frac{A_k}{4 k} \left( 1 \pm \sqrt{1-4 \frac{m_N^2}{A_k^2}}
 \right) \,.
\label{phic}
\eeq
Depending on the sign of the mass squared $m_{\phi}^2$,
$\phi$ can roll towards $\phi^+$ from above ($m_{\phi}^2>0$, hybrid inflation)
or towards $\phi^-$ from below ($m_{\phi}^2<0$, inverted hybrid
inflation \cite{inverted}). Either way, when the critical value is
reached, inflation ends and the following global minimum is achieved:
\begin{eqnarray}
   <\phi>&=& \frac{A_k}{4 k} \,, \label{phivev}\\
    <N>  &=& \frac{A_k}{2\sqrt{2} k}\sqrt{1-4 \frac{m_N^2}{A_k^2}}
=\sqrt{2} \left|
   \phi_c^{\pm}-<\phi > \right| \label{nvev}\,,
\end{eqnarray}
During inflation the potential energy
of the vacuum takes the positive value $V(0)$ (which will be shortly
explained in the context of a SUGRA model) which drives inflation.
After inflation ends $V(0)$ is assumed to remain unchanged,
but be cancelled by a negative contribution from the remaining part
of the potential at the global minimum:
\begin{equation}
V(<\phi >,<N>)= -k^2<N>^4=-4 k^2 (\phi_c^{\pm} - <\phi >)^4 .
\label{V0}
\end{equation}
The cancellation mechanism is beyond the scope of this paper
(this is the problem of the cosmological constant).

We now wish to elevate this model to an SUGRA theory, using the
results of earlier sections, and in particular the Kahler potential in
Eq. (\ref{Kfinal}). The inflaton $\phi$ is in the untwisted sector, and
appears with the overall modulus T in the combination $\rho$. 
As before $T$ is assumed not to
enter the superpotential. The singlet $N$ and the remaining MSSM
superfields are in the twisted sector with 
canonical Kahler potentials (zero modular weights) to begin with.
The  vacuum energy driven inflation $V(\rho_0,S_0)$ is achieved by the partial
cancellation between the $\rho$ and the dilaton contributions. 
The superpotential $W$ is from Eq. (\ref{W}),
\beq
W=-kN^2\phi+\hat{W}(S) \,,
\eeq
where $\hat{W}(S)$ was given in Eq. (\ref{WS}).
In the present SUGRA framework we can compute the soft parameters
$m_N$ and $A_k$, and the related  critical value  $\phi_c^{\pm}$ and
vevs of the fields, in term of the gravitino mass. We are now
interested in what happens after inflation to the fields $\phi$ and
$N$, so first we have to reincluded them in the SUGRA
potential. The potential is then that of Eq. (\ref{Vrhos}) plus the
contribution of $\phi$ and $N$:
\beq
V=V(\rho,S)+e^K\left[-\frac{|W_{\phi}|^2}{K'} + |N^\ast W + W_N|^2
\right] \,.
\label{Vnphi1}
\eeq
 When inflation ends, we can expect the value of $\rho$ and $S$ to
change, as all the  
fields will now adjust their values in the global
minimum of the theory. Nevertheless, as far as the contribution to
the superpotential due to fields $N$ and $\phi$ is suppressed respect
to the dilaton term, the minima for the fields $S$ or $\rho$ (and now
by extension also $T$) will be only slightly shifted, and they will
safely remain where they were previously. 
The fact that the minima for $\rho$ and $S$ is almost unchanged after
inflation also relies on the condition $\epsilon \ll 1$. This 
together with $\tilde{W}/\hat{W} \sim O(\epsilon)$ makes the 
minimization conditions for $\rho$ and $S$ to look the same than
those given in Eqs. (\ref{minrho}) and (\ref{minS}), and at most the
values of $\rho_0$ and $S_0$ will be shifted only by a factor
$O(\epsilon)$. Therefore the gravitino mass
will also take the same value during and after inflation, i.e.,
\beq
m_{3/2}^2 = <e^G> \simeq e^K \frac{|\hat{W}(S)|^2}{\tilde{M}_P^4} \simeq
\frac{\Lambda^6} {\tilde{M}_P^4} \,.
\eeq
In order to keep a gravitino mass of order $1\, TeV$ we require
$\Lambda \approx 10^{13}\,GeV$. 

In the limit of large moduli, we
can now expand the potential in powers of $\phi \phi^\ast/(T+T^\ast)$.    
The remaining explicit dependence on the modulus $T$  will be absorbed in the
proper normalisation of fields and Yukawas to get the physical degrees
of freedom and couplings, and also in the gravitino mass. Therefore we
can write our potential as, 
\beq
V=V(\rho_0,S_0)+|W_{\phi}|^2  + |W_N|^2 +
m_{3/2}^2 |N|^2 + m_{3/2} ( N W_N + h.c.) -\delta ( \tilde{W}(\phi,N) + h.c.)
+ \ldots\,, 
\label{Vnphi2}
\eeq
where the dots stand for non-renormalisable terms, and from
Eq. (\ref{Vep}) $V(\rho_0,S_0)\simeq
\epsilon m_{3/2}^2 \tilde{M}_P^2$ . According to
Eq. (\ref{Vnphi2}) the soft parameters are defined as, 
\bea
m^2_N &\equiv& m^2_{3/2} \label{mn} \,, \\
A_k   &\equiv& (2-\delta) m_{3/2} < 2 m_{3/2} \label{Ak}\,, \\
\delta &=& 3 - \frac{K'^2}{K''} + \frac{K_S}{b_0 K_{SS}}\,.
\eea
It is clear from Eqs. (\ref{nvev}),
(\ref{mn}), (\ref{Ak}) that in this particular case\footnote{We will
see later that in this model the coupling between $\phi$ and $N$ is too
small, and therefore radiative corrections are not enough to break the
relation $A_k^2<4 m^2_N$ to the extend needed.} no minima exists
different than $<N>=0$, and thus $V(<\phi >,<N>)= 0$. 

In order to avoid this problem we allow the field $N$ to have a non-zero
modular weight $n_N$. An interesting possibility is to have $N$ in the
untwisted sector, on the same foot in the Kahler potential than the
field $\phi$. That is, $N$ will enter into the combination $\rho$,
will be massless, 
but still  $N=0$ during inflation. The Kahler potential and the
potential in this case becomes: 
\bea
K&=& -3 \ln \rho +\frac{\beta}{\rho^3} -\ln (S+S^\ast)+\hat{K}_{np}(S)
\,,\\
 \rho &=&T+T^\ast - \phi^\ast \phi - N^\ast N \,, \\
V &=& e^K \left[ (\frac{K'^2}{K''}-3 ) |W|^2
  +\frac{1}{K_{SS}} | K_S W + W_S |^2  
  -\frac{|W_{\phi}|^2}{K'}-\frac{|W_{N}|^2}{K'}
\right]  \nonumber \\
  &=& V(\rho_0,S_0)+|W_{\phi}|^2  + |W_N|^2
 - \delta m_{3/2} ( \tilde{W}(\phi,N) + h.c.) +\ldots 
\,,
\label{Vnphi3}
\eea
with the soft parameters given now by\footnote{We implicitly assume
that the sign of the $k \phi N^2$ coupling in the superpotential may
be changed in order to match Eq.(\ref{Vnphi3}) with Eq. (\ref{Vnphi}).},
\bea
m^2_N &=& 0\,, \\
A_k   &=&  \delta m_{3/2} \label{Ak3}\,,
\eea
and the condition $A_k^2 > m_N^2$ trivially satisfied. 

The main point to note is that 
the soft parameters are of order $m_{3/2}\sim 1$ TeV both during and after
inflation, but the inflaton during inflation will have a zero
tree-level mass due to the no-scale mechanism.
 After inflation ends
the value of $V(\rho_0,S_0)$ is assumed to be unchanged, but there is
in general a negative contribution from the remaining part of the
potential in Eq. (\ref{V0}) whose value depends on the coupling $k$,
which in this model is fixed to be very small, as we now discuss.

One of the additional features of the NMSSM of hybrid inflation is
that it can solve the strong CP problem due to the presence of an
$U(1)_{PQ}$ symmetry in the superpotential given in
Eq. (\ref{WNMSSM}), broken by the vevs of $\phi$ and $N$. This implies
that there is an axion field, and to satisfy the cosmological
constraints on its abundance it is required $<N> \sim <\phi> \sim
10^{13}\, GeV$. Looking at Eqs. (\ref{phivev}-\ref{nvev}) we end with the
condition $k \sim 10^{-10}$, a quite small coupling constant, but that
will render the value of the potential in Eq. (\ref{V0}) of order, 
\beq
V(<\phi >,<N>) \sim -( 10^8 \, GeV )^4\,,
\eeq
This is in fact the same order as the potential $V(\rho_0,S_0)
\simeq \epsilon m^2_{3/2} \tilde{M}^2_P$
obtained in the previous section when $\epsilon \simeq 10^{-10}$. Therefore we may propose that at the 
global minimum these two potentials of opposite sign 
and the same magnitude accurately cancel
to yield an acceptably small cosmological constant.
With the extremely small coupling $k$ radiative corrections yield an
(ultralight) inflaton mass during inflation in the eV range
\cite{NMSSM}, assuming that the no-scale mechanism 
developed in this paper sets its tree level
mass during inflation to zero. As discussed in detail elsewhere \cite{NMSSM}
an inflaton mass in the eV range together with a potential during inflation
of height $V(0)^{1/4} \sim 10^8 \, GeV $ gives an appropriate COBE
normalisation.

We end this section by noting that
the smallness of $k$ seems to indicate that in fact the original
coupling has a non-renormalisable origin, as previously discussed \cite{NMSSM},
with $k$ effectively given as a ratio of two
scales, namely $k \sim (\Lambda/ \tilde{M}_P)^2$,
where $\Lambda$ was identified as the VEV of some new fields.
In the present framework we might try to 
identify the scale $\Lambda$ with the effective scale in the dilaton
superpotential, since these two scales
are naturally the same order of magnitude in this model.
However we shall not pursue this possibility further here.

\section{Final Comments}

We should comment on the effect of strongly coupled
heterotic string theories on models
such as this which rely on no-scale SUGRA. 
In general one might expect that the overall modulus field $T$
will appear in combination with the dilaton $S$ in the combination
$S+\alpha T$ in the dilaton superpotential, thereby invalidating the
no-scale assumption that $T$ does not enter the superpotential,
and so invalidating our basic assumption that the inflaton field $\phi$
is massless at the string scale. However the situation is in fact not
so clear since some authors claim that the parameter $\alpha$ may in fact
be extremely small even in strongly coupled heterotic string (M) theory.
The result may depend on the precise mechanism for SUSY breaking in the 
hidden sector (e.g. gaugino condensation vs. Scherk-Schwartz breaking)
\cite{SS}. In any case the discussion of such issues, along with the
dilaton potential, lies beyond the scope of the present paper.

In this paper we have shown that effective SUGRA theories
of the no-scale form which may arise naturally from string theory
may lead to viable models of F-term hybrid inflation.
Whilst it is true that perturbative string theory cannot explain dilaton
stabilisation, it is generally accepted that non-perturbative
string theory could provide a resolution to this problem in the
future, and it has become commonplace to introduce {\it ad hoc}
Kahler potentials to model such an unknown non-perturbative
behaviour of string theory. In this spirit we introduced
the Kahler potential in Eqs. (\ref{Kdilanp}), (\ref{betabit})
and the superpotential in Eq. (\ref{WS})
to model the unknown non-perturbative effects of string theory.

The main point of this paper is to construct a general framework
for pursuing $F$ term inflation in effective SUGRA theories which
may typically arise in string models. As we have discussed, such
theories are typically of the no-scale form, and we have proposed that
it is natural to place the inflaton in the untwisted sector
along with the moduli fields in order to obtain a massless inflaton
and so solve the $\eta$ or slow roll problem. To summarise, we have
assumed the following conditions to hold during inflation:

(a) The superpotential is independent of the modulus $T$.

(b) The Kahler potential depends on the combination $\rho= T+T^\ast
-\sum_i \phi^\ast_i \phi_i$, with $\phi_i$ any untwisted field present in the
theory. Here we have only consider those relevant for
hybrid inflation, i.e, $\phi$ and $N$. In addition, there are non-perturbative
contributions for both the field $\rho$ and the dilaton $S$ which help
to stabilise them at a fixed value during inflation. 

(c) The SUGRA potential during inflation is of the form:
\beq
V= |F_T|^2 + |F_S|^2 - 3 e^G\,,
\eeq
such that the positive contribution can be fine-tuned to cancel the
negative term to the extend needed to  generate a positive potential
energy suitable for inflation. 

The first two conditions ensures the masslessness of the inflaton. The
dilaton role is twofold: it works as a source for SUSY breaking, and
its positive contribution to the potential helps to keep the
potential scale during inflation below the typical SUSY breaking
scale.
  
We emphasise that it is not necessary for the theory
to possess a Heisenberg symmetry, but only that the other
twisted  fields with modular weights $n_i<-1$ be switched off during
inflation. In such theories  
the partial cancellation of the negative term in the SUGRA potential
against the dilaton and moduli contribution 
allows the possibility that the height of the potential
during inflation is much smaller than usually assumed, 
 with $V \sim \epsilon m_{3/2}^{2} \tilde{M}_P^{2}$, the size of
$\epsilon$ being model dependent. 
 Smaller potentials are welcome in this
kind of approach because one must not only ensure that the inflaton
is massless during inflation at tree level, but also explain why
radiative corrections do not lead to the inflaton acquiring a 
mass which would be again  too large to avoid the $\eta$ problem.
In order to control the radiative corrections we have simply assumed
that they  are controlled by very small couplings.
The presence of such small couplings is then generally associated with
small height potentials which are therefore a generic consequence of our
approach. Another good feature of having small couplings in the
hybrid superpotential is that their contribution when inflation ends
will be highly suppresed respect to the dominant dilaton contribution,
and as a result the minima for the moduli and the dilaton are mainly
the same during and after inflation. That is, the generic cosmological
moduli problem \cite{moduliproblem} will not be present in our
scenario with a small height potential.

To illustrate these features we have revisited an NMSSM
model of inflation corresponding to the superpotential in Eq. (\ref{WNMSSM}) 
in which the radiative corrections
are controlled by an extremely small Yukawa coupling $k\sim 10^{-10}$ which
leads to an ultralight inflation in the eV range. Such a light inflaton,
combined with the COBE normalisation requires a potential during inflation of
$10^{8}$ GeV which corresponds to Eq. (\ref{Vep}) with
$\epsilon^{1/4} \sim 10^{-3}$.
Of course such small couplings may be regarded a rather extreme example
of how to control the radiative corrections, but it is worth recalling
that the strength of the couplings
are determined by the requirements of satisfying the COBE normalisation
solving the strong CP problem, and the $\mu$ problem at the same time
\cite{NMSSM}. It is also worth recalling that the
origin of such small couplings may involve an
additional sector which obeys a discrete $Z_3\times Z_5$ symmetry
from which the Peccei-Quinn symmetry emerges as an approximation.
We have not performed an explicit string compactification
to generate the desired symmetry, nor have we derived the
superpotential from first principles
(which would involve standard techniques of taking
vacuum expectations of vertex operators). However
explicit string constructions do exist in the literature
which involve similar discrete symmetries to those assumed here,
and which forbid the lowest
dimension operators but allow operators at higher order.

Finally although it might seem that the present model is quite
complicated, it is worth emphasising that any realistic supergravity model
of inflation will necessarily involve a similar number of free parameters.
The literature is replete with toy models or incomplete theories
which look simpler than our model precisely because they are not
complete. Our model, although not derived from first principles
using string theory, nevertheless represents a complete working model
of no-scale supergravity inflation. It contains explicit mechanisms
for dilaton and moduli stabilisation and supersymmetry breaking,
and it successfully resolves the $\eta$ problem of supergravity
hybrid inflation. It looks more complicated precisely because
it goes further towards being fully realistic than most other
supergravity models. We have already emphasised that the
model solves the strong CP problem and the $\mu$ problem
of supersymmetry. The model clearly has many virtues, and
represents the most realistic attempt at supergravity
hybrid inflation (as far as we are aware) available in the
current literature. It remains to be seen if its central
prediction, a spectral index $n$ precisely equal to unity \cite{NMSSM},
will be verified by the forthcoming Map and Planck explorer experiments.

\begin{center}
{\bf Acknowledgements}
\end{center}
We would like to thank Emilian Dudas, David Lyth and
Toni Riotto for useful comments.

\end{document}